\pdfoutput=1
\documentclass[aps,prd,tightenlines,floatfix,twocolumn,superscriptaddress,showpacs]{revtex4}

\usepackage{graphicx}
\usepackage{wrapfig}
\usepackage{hyperref}
\usepackage[all]{xy}
\usepackage[usenames]{color}

\newcommand{\nc}{\newcommand}
\nc{\beq}{\begin{equation}}
\nc{\eeq}{\end{equation}}
\nc{\bea}{\begin{eqnarray}}
\nc{\eea}{\end{eqnarray}}
\nc{\n}{\nonumber \\}

\nc {\araa}{Annual Review of Astronomy and Astrophysics}
\nc{\physrep}{Physics Reports}
\nc{\apjl}{Astroph. J. Letters}

\nc{\tpm}{\tau^+\tau^-}
\nc{\bb}{b \bar b}
\nc{\ee}{e^+e^-}
\nc{\mm}{\mu^+\mu^-}
\nc{\ww}{W^+W^-}
\nc{\acs}{\ensuremath{ \langle \sigma_{\rm a} v \rangle  }}

\nc{\MLsun}{\ensuremath{M_\odot/L_\odot}}
\nc{\seg}{Segue~I}
\nc{\pone}{Paper~I}
\nc{\ptwo}{Paper~II}

\begin{document}  

\date{\today}
\title{Green Bank Telescope Constraints on Dark Matter Annihilation in Segue~I}

\author{Aravind Natarajan}
\email{anat01@me.com}
\affiliation{Department of Physics, Engineering Physics, and Astronomy, Queen's University, Kingston, Ontario K7L 3N6, Canada}
\affiliation{Department of Physics and Astronomy, University of Pennsylvania, 209 South 33rd Street, Philadelphia, PA 19104, USA}

\author{James E. Aguirre}
\affiliation{Department of Physics and Astronomy, University of Pennsylvania, 209 South 33rd Street, Philadelphia, PA 19104, USA}

\author{Kristine Spekkens}
\affiliation{Department of Physics, Royal Military College of Canada,
PO Box 17000, Station Forces, Kingston, Ontario, Canada K7K 7B4}

\author{Brian S. Mason}
\affiliation{National Radio Astronomy Observatory, 520 Edgemont Road, Charlottesville, VA 22903-2475, USA}

\begin{abstract}
We use a non-detection in $\nu = 1.4\,$GHz Green Bank Telescope observations of the ultra-faint dwarf spheroidal galaxy \seg, which could be immersed in a non-negligible halo magnetic field of the Milky Way, to place bounds on particle dark matter properties. We model the galaxy using an Einasto dark matter profile, and compute the expected synchrotron flux from dark matter annihilation as a function of the magnetic field strength $B$, diffusion coefficient $D_0$, and particle mass $m_\chi$ for different annihilation channels.  The data strongly disfavor annihilations to $e^+e^-$ for $m_\chi \lesssim 50\,$GeV, but are not sensitive to the $b \bar b$ channel.  Adopting a fiducial $B \sim 2\,\mu$G inferred from \seg's proximity to the Milky Way, our models of annihilation to $\tau^+\tau^-$ with $m_\chi = 30\,$GeV require an intermediate value of $D_0$ for consistency with the data. The most compelling limits are obtained for WIMP annihilation to $\mu^+\mu^-$: we exclude $m_\chi \lesssim 30\,$GeV$\,\rightarrow\mu^+\mu^-$ at 95\% confidence, unless $D_0$ exceeds the Milky Way value or $B$ is significantly smaller than we have assumed.


\end{abstract}




\maketitle

\section{Introduction}

 Weakly Interacting Massive Particles (WIMPs) are one of the leading candidates for the dark matter of the Universe, and a huge world-wide effort is underway to detect them through direct, indirect, and collider experiments. WIMPs are natural dark matter candidates because they were put forward to solve problems in particle physics unrelated to the dark matter puzzle. They also predict the correct relic density at the present epoch almost independently of the mass. More importantly, the presence of weak interactions allows us to make testable predictions.

Indirect dark matter detection experiments include gamma ray observations of the Milky Way center and dwarf galaxies \cite{ackermann_etal_for_fermi, fermi_new},  cosmic microwave background measurements \cite{cmb1, cmb2, cmb3, cmb4, cmb_arvi}, Cherenkov observations \cite{cherenkov} and neutrino flux measurements \cite{neu1,neu2}.  Interestingly, observations of the Milky Way center \cite{gal_center} by the Fermi  gamma ray telescope seem to indicate an excess of gamma rays consistent with dark matter particles of mass $m_\chi = 31-40$ GeV annihilating  at a rate $\acs = (1.4 - 2.0) \times 10^{-26}$ cm$^3/$s, close to the thermal relic value. More recently, authors \cite{ret1,ret2} found an excess of gamma rays from the dwarf galaxy Reticulum~II, possibly consistent with dark matter annihilation. Models favour very light WIMPs for annihilation into $\tau^+\tau^-$ and $\mu^+\mu^-$ \cite{ret1}, with most likely values of $m_\chi \sim 15\,$GeV  and  $m_\chi \lesssim10 \,$GeV for these channels respectively.

In this article, we explore dark matter detection through synchrotron radiation from dwarf galaxies. Particle annihilation results in the release of high energy charged particles which emit synchrotron radiation in a magnetic field. The detection or absence of this synchrotron radiation can be used to place useful bounds on dark matter properties \cite{cpu1,cpu2}.

\begin{figure*}
\scalebox{0.4}{\includegraphics{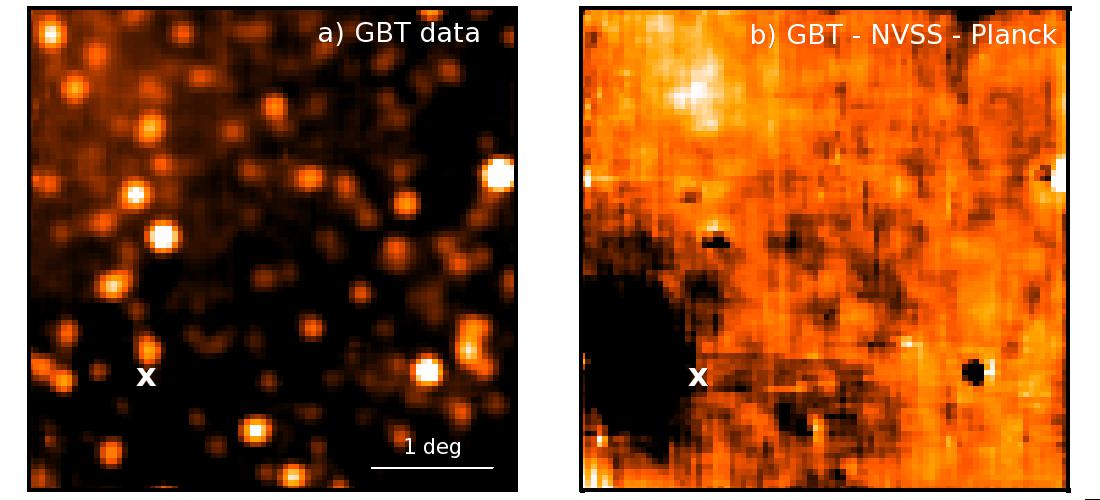}}
\caption{GBT $\nu=1.4\,$GHz radio continuum maps in the \seg\ region. Plot (a) shows the calibrated GBT map, which includes background point sources and diffuse sky emission. The linear radiation specific intensity scale ranges from -10 to 250$\,$mJy/beam. Plot (b) shows the mean-subtracted residual map, with point sources removed using the NVSS catalog and diffuse sky removed using Planck maps. The linear radiation specific intensity scale ranges from -15 to 30$\,$mJy/beam. The horizontal line in plot (a) is 1 degree in length, and applies to both panels. The cross in both plots denotes the optical centroid of \seg, and the symbol size resembles the 9~arcmin elliptical half-light diameter of the dwarf. We find no excess radio emission at the location of \seg\ in the mean-subtracted residual map.
\label{fig1} }
\end{figure*}

 Dwarf spheroidal galaxies are interesting targets for synchrotron radiation searches from annihilating dark matter because of their proximity and their inferred high dark matter content \cite{mateo,strigari_etal_2007, strigari_etal_2008, strigari_etal_nature_2008}.  In particular, observed line-of-sight velocities of individual stars belonging to the ultra-faint dwarfs suggest that they have significantly higher mass-to-light ratios within their half-radii than the classical dwarfs \cite{Geha09}, making them the most likely places to observe an electromagnetic signature of annihilating dark matter.  Some of the extreme ultra-faint dwarfs ($L \lesssim 10^3 L_{\odot}$) have a well-measured velocity dispersion, and the appearance of being in dynamical equilibrium \cite{simongeha07, kirby13, walker13}.
 
Several radio synchrotron searches for annihilating dark matter signatures in dwarf spheroidal galaxies have been carried out. In previous work \cite[][hereafter \pone]{paper1}, our group obtained deep $\nu=1.4\,$GHz radio observations of Draco, Ursa Major II, Coma Berenices, and Willman I using the Robert C. Byrd Green Bank Telescope (GBT), and used the resulting maps of Ursa Major II and Willman I to constrain models considered by \cite{cpu2}. A more detailed analysis of dark matter annihilation in Ursa Major II was described in \cite[][hereafter \ptwo]{arvi_dwarf}.  More recently, authors \cite{atca1, regis_etal, atca3} used the Australia Telescope Compact Array (ATCA) to search for radio emission from the dwarf galaxies visible from the southern hemisphere. These observations place tighter constraints than in \pone\ and \ptwo\ for compact annihilation signatures, but are not sensitive to the radio emission on scales $\gtrsim 15\,$arcmin produced by many annihilation channels. 
 
Synchrotron radiation searches for annihilating dark matter hinge on the magnetic field strength $B$ of the source, but not much is known about $B$ in dwarf spheroidal galaxies \cite{beck13}. The previous work described above relied on scaling arguments from detections in dwarf irregular galaxies \cite{chyzy} or optimistic extrapolations of the Milky Way's vertical field  \cite{carretti_etal} in adopting  $B \sim 1\mu$G. However, our recent search for atomic gas in dwarf spheroidal galaxies \cite{spekkens14} suggests that they have a more tenuous interstellar medium than expected from stellar evolution models. This raises the possibility that $B$ is much lower than the fiducial value previously invoked, producing correspondingly weaker particle dark matter constraints.

 \seg\ (RA = 10:07:03.2 $\pm$ 1.7$^s$, DEC = +16:04:25 $\pm$ 15'') \cite{segue1} is an ultra-faint dwarf spheroidal galaxy at a distance $\sim$ 23 kpc from the sun, with a mass-to-light ratio $M/L \sim3400 \,$\MLsun\ \cite{Geha09,segue3}, making it one of the darkest galaxies known so far. The large mass-to-light ratio combined with the close proximity of \seg\ make it an excellent target for indirect dark matter searches. 
 More importantly, \seg\ is among the closest dwarf spheroidal galaxies to the sun, which means that it could be immersed in the outer halo magnetic field of the Milky Way itself \cite{regis_etal}. We therefore expect a non-negligible ambient magnetic field at the location of \seg, which makes it a unique target for synchrotron radiation searches for annihilating dark matter. In this article, we make use of new GBT observations of \seg\ to carry out this search. 


\section{GBT observations and analysis of Segue I}
\label{data}

Observations sensitive to degree-scale radio emission from dwarf spheroidal galaxies require the use of a single dish telescope. On the other hand, the low angular resolution of single dishes means that we require an interferometer to identify and subtract background point sources. The GBT is a fully steerable 100m single dish antenna in West Virginia that is well-suited to study radio emission in the 300 MHz - 100 GHz frequency range. In order to search for extended radio emission from \seg\, we used the GBT to map a $4^\circ \times 4^\circ$ square region in its vicinity at an observation frequency $\nu$ = 1.4 GHz, affording the use of publicly available, 45 arcsecond resolution NRAO VLA Sky Survey (NVSS) data \cite{nvss} at that same frequency to carry out the point source subtraction. 

\begin{figure*}[t]
\scalebox{0.7}{\includegraphics{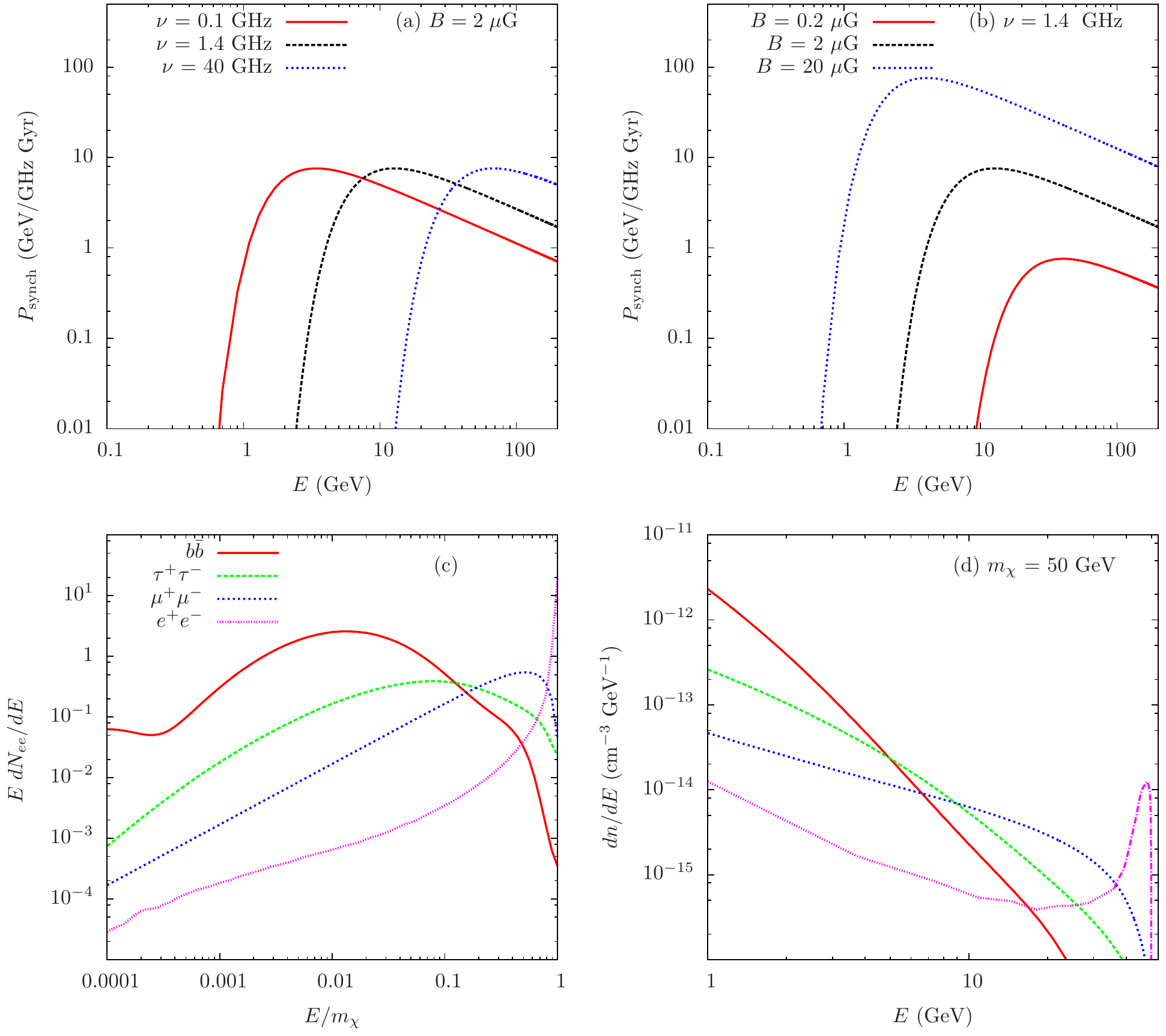}}
\caption{ Synchrotron radiation from dark matter annihilation. Plots (a) and (b) shows the synchrotron kernel $P_{\rm synch}$ as a function of observation frequency $\nu$ and magnetic field strength $B$, respectively. The sharp cut-off at low particle energies $E_{\rm cutoff}$ implies a similarly sharp cutoff at small WIMP masses $m_\chi < E_{\rm cutoff}$. Plot (c) shows the particle spectrum from dark matter annihilation for different annihilation channels, and plot (d) shows the particle number density density per unit energy after transport at $r$ = 0.1 kpc, for $m_\chi$ = 50 GeV, $B$ = 2 $\mu$G, and $D_0$ = 1 kpc$^2/$Gyr for those same annihilation channels. The synchrotron radiation at a given observing frequency is sensitive to $B$, $m_\chi$ and the annihilation channel. 
\label{fig2} }
\end{figure*}

The GBT observations of \seg\ were obtained in on-the-fly mode during several sessions in late 2013 under program AGBT13B253 using the same setup as described in detail in \pone. The maps are calibrated, point-source and diffuse-sky subtracted in a single step by building models of the point source and diffuse sky from the NVSS catalog and the Planck 30 GHz synchrotron model \cite{planck15}.  Both NVSS and Planck templates are convolved with the GBT beam and simultaneously fit to the GBT map, with a single amplitude for each template map.  The amplitude of the fit determines the calibration of the GBT maps into radiation specific intensity units.


We use an improved mapping technique to recover more extended radio emission (and thus a higher sensitivity to dark matter annihilation signals) in the final maps than achieved in \pone. Single dish data, particularly continuum data, have time variable mean levels due to varying ground spillover, changing atmospheric emission, and low-level electronic gain fluctuations. These temporal baselines were removed in \pone\ by fitting simple linear functions to each scan, which had the side-effect of suppressing extended structure in the map. The effect is illustrated in fig.\ 7 of \ptwo, where as much as 85\% of the extended map flux is filtered out. We have therefore implemented a de-striping calibration algorithm \cite{haslam70,haslam81,winkel12} that uses only the data differences to solve for the baseline terms, resulting in full sensitivity to diffuse sky structure. 

Fig.~\ref{fig1} shows the calibrated GBT maps of \seg\  in (RA, DEC) coordinates in standard radiation specific intensity units of mJy per 9.1 arcmin telescope beam full-width at half-maximum diameter (mJy/beam). The horizontal line in plot (a) is 1 degree in length, and applies to both panels. The crosses in both panels show the stellar centroid location of \seg, with the symbol size reflecting the $\sim$9~arcmin elliptical half-light diameter of the dwarf. Plot (a) shows the calibrated GBT map, which includes background point sources and diffuse sky emission.  The latter is most evident in the top-left corner of Fig.~\ref{fig1}, where a 1.5-deg ``warm spot" clearly underlies the background point sources. 

\begin{figure*}[!t]
\scalebox{0.32}{\includegraphics{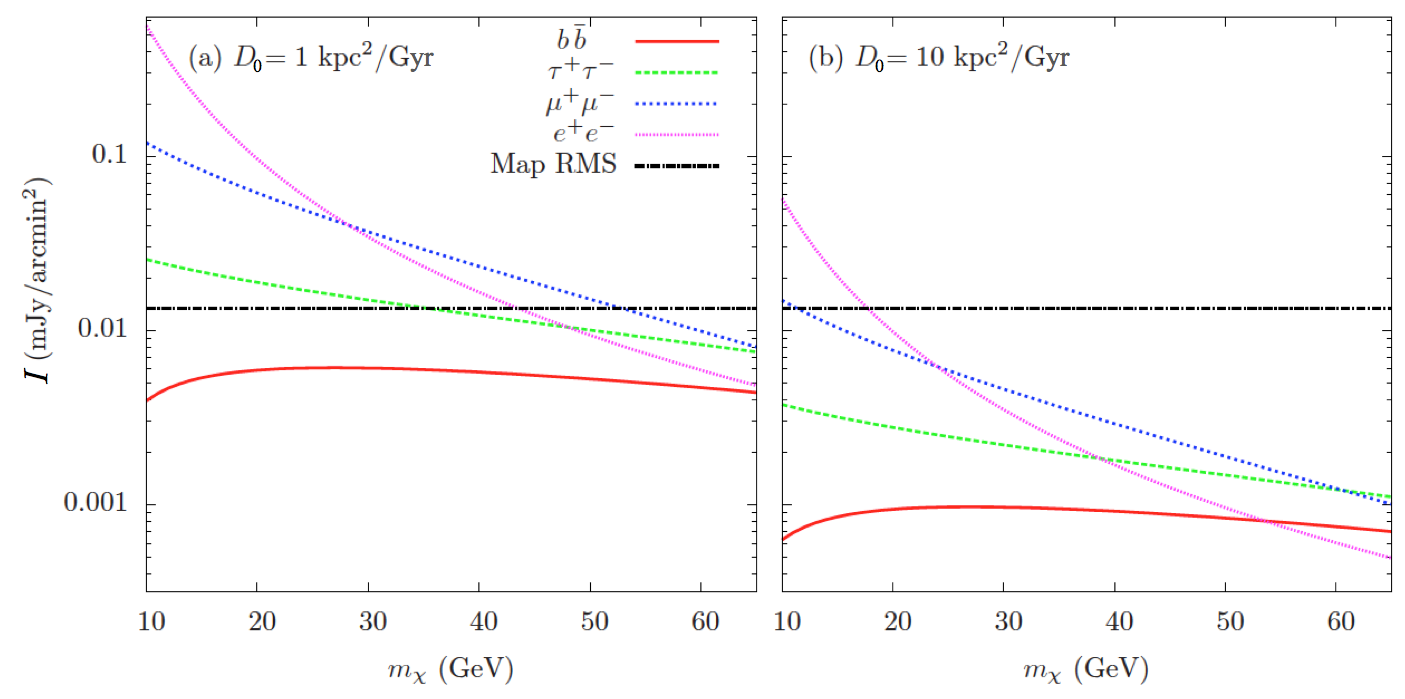}}
\caption{Predicted $\nu=1.4\,$GHz radio emission from dark matter annihilation in \seg\ for Bin \#1 ($0^\circ < \theta < 1.12^\circ$) for different annihilation channels, as a function of the WIMP mass $m_\chi$. All models adopt a fiducial magnetic field $B = 2\,\mu$G. Plot (a) shows models with a diffusion coefficient $D_0 = 1\,$kpc$^2/$Gyr, and plot (b) shows the same models but with $D_0=10\,$kpc$^2/$Gyr.  The dashed horizontal line in both plots shows the RMS fluctuation for Bin \#1 in the mean-subtracted residual map in Fig.~\ref{fig1}b, obtained from simulated maps with the same power spectrum and variance as the data. The $b \bar b$ channel is undetectable in our data, whereas our non-detection in Bin \#1 strongly rules out light WIMPs annihilating into $e^+e^-$.
\label{fig3} }
\end{figure*}

Plot (b) of Fig.~\ref{fig1} shows the mean-subtracted residual map, now on a much narrower linear intensity scale than in plot (a), after background point-source removal using the NVSS data and diffuse sky emission removal using the Planck map. Some point-like residuals remain in the vicinity of the brighter sources; this could indicate variability in the $\sim$20-year  time span between the NVSS and GBT data acquisition, or could arise from small astrometric differences between these datasets. These artifacts are on a much smaller angular scale than the dark matter annihilation signatures we are searching for, and are therefore unimportant in our analysis. There are also some larger-scale features in the map in Fig.~\ref{fig1}b, such as the cold spot to the West of \seg\ and the warm spot to the North. They likely represent residual Milky Way foreground emission that was not adequately subtracted with the scaled Planck map. This is not surprising, since small spectral index variations in these features between the $\nu=30\,$GHz Planck observation frequency and our $\nu=1.4\,$GHz observations would lead to improper subtraction using our adopted technique. As in \pone, these foregrounds limit the ultimate map sensitivity. 

Fig.~\ref{fig1}b shows that the optical centroid of \seg\ sits on the edge of a cold spot in the mean-subtracted residual map. In particular, we find no residual emission that spatially correlates with \seg. We therefore proceed to compare this non-detection to the synchrotron radiation signatures expected for different dark matter annihilation channels.

\begin{table*}
\caption{Binned data and model radiation specific intensities}
\centering
\begin{tabular}{|c|c|cccc|cccc|}
\hline
\hline 
                                             &                                & \multicolumn{4}{c|}{$\mathcal{I}_{model}$, $D_0 = 1\,$kpc$^2/$Gyr} & \multicolumn{4}{c|}{$\mathcal{I}_{model}$, $D_0 = 10\,$kpc$^2/$Gyr} \\
             Bin                          &         $\mathcal{I}_{data}$               &  \multicolumn{2}{c}{$m_\chi=10\,$GeV} & \multicolumn{2}{c|}{$m_\chi=30\,$GeV} &   \multicolumn{2}{c}{$m_\chi=10\,$GeV} & \multicolumn{2}{c|}{$m_\chi=30\,$GeV} \\
                                            &                                & $\tau^+\tau^-$ &$\mu^+\mu^-$&$\tau^+\tau^-$&$\mu^+\mu^-$&$\tau^+\tau^-$&$\mu^+\mu^-$&$\tau^+\tau^-$&$\mu^+\mu^-$\\
 \hline
 \#1 (mJy/arcmin$^2$) & $0.010 \pm 0.013$ &  0.026 & 0.119 & 0.015 & 0.037 & 0.0037 & 0.015 & 0.0022  & 0.0046\\
 \#2 (mJy/arcmin$^2$) & $0.018 \pm 0.017$ & 0.0025 & -0.0089 & 0.0002 & 0.00015 & 0.0037 & -0.0017  & 0.000016 & 0.00003\\
 \#3 (mJy/arcmin$^2$) & $-0.0097 \pm 0.017$ & -0.010 & -0.049 & -0.0061 & -0.0152 & -0.0015 &  -0.0067 & -0.00091& -0.0021\\
 \#4 (mJy/arcmin$^2$) & $0.0029 \pm 0.013$ & -0.013 & -0.062 & -0.0081 & -0.022 & -0.0024 & -0.0064 & -0.0015 & -0.0023\\
    \hline
\end{tabular}
\label{t1} 
\end{table*}

\section {Synchrotron Radiation from Dark Matter Particle Annihilation}
\label{models}

We model the synchrotron radiation from dark matter particle annihilation in \seg\ in a similar manner as in \ptwo.
The transport of charged particles is affected by diffusion and energy losses. Under the simplifying assumptions of spherical symmetry and a constant magnetic field, we may solve for the number of electrons/positrons per unit energy per unit volume $dn/dE$ at a distance $r$ from the center of the dwarf \cite{cpu1,cpu2}:
\beq
D \; \nabla^2 \left( \frac{dn}{dE} \right )  + \frac{\partial}{\partial E} \left( b \frac{dn}{dE} \right ) + Q = 0 \,\,\,.
\label{diffusion}
\eeq
The source term $Q$ is assumed to be entirely due to dark matter annihilation:
\beq
Q(r,E) = \frac{\langle \sigma_{\rm a} v \rangle}{m^2_\chi} \, \rho^2_\chi \, \frac{dN}{dE} \,\,\,,
\label{Q}
\eeq
where $\langle\sigma_{\rm a} v \rangle$ is the annihilation rate assumed constant at the thermal value in all our models \cite{arvi_mssm}:
\beq
\acs = 2.2 \times 10^{-26} \; {\rm cm}^3/{\rm s} \,\,\,,
\eeq

and $dN/dE$ is the number of $e^+e^-$ particles produced per WIMP:
\beq
\int dE \, E \, \frac{dN}{dE}  \leq 1 \,\,\,.
\eeq

Note that we have chosen $dN/dE$ to be the particle spectrum per WIMP, and not per annihilation (otherwise there would be an extra factor of 2 in the denominator of Eq.~\ref{Q}). The form of $dN/dE$ depends on the annihilation channel. For direct annihilation of WIMPs to $e^+e^-$, $dN/dE$ resembles a delta function at the WIMP mass. For other channels, $dN/dE$ is a continuous function of energy. 

The dark matter density profile $\rho_\chi$ of \seg\ is assumed to follow the Einasto form:
\beq
\rho_\chi(r) = \rho_{\rm s} \exp \left\{ -2\alpha \left[ \left(\frac{r}{r_{\rm s}} \right )^{1/\alpha} - 1 \right ] \right\} \,\,\,.
\eeq
We choose $\alpha = 0.3$ to obtain an inner density slope $\rho \sim r^{-0.3}$ in agreement with The HI Nearby Galaxy Survey (THINGS) mass models \cite{things1,things2,things3}. We choose $r_{\rm s}$ = 0.15 kpc \cite{magic1, essig_etal_2009} and $\rho_s$ = 6.6 GeV/cm$^3$ in order to reproduce the Fermi computation \cite{ackermann_etal_for_fermi} of the emission measure for \seg:
\bea
J = \int d\Omega \; \int ds \, \rho^2_\chi(r) \n
\log_{10} \left( \frac{J} {  {\rm Gev}^2 {\rm cm}^5 } \right) = 19.6 \,\,\,,
\eea
where the integral over solid angle is from 0 to 0.5 degrees, $s$ is the distance measured along the line of sight, and the assumed distance to \seg\ is $L$ = 23 kpc.

In Eq.~\ref{diffusion}, $D$ is the diffusion parameter which takes the form:
\beq
D(E) = D_0 \left( \frac{E}{E_0} \right )^\gamma \,\,\,,
\eeq
where $D_0$ is the diffusion coefficient, and the index $\gamma=0.7$ in accordance with the median Milky Way value \cite{donato_etal_2004}. $b$ is the energy loss term \cite{cpu1}:
\beq
b(E) = b_0 \left ( \frac{E}{E_0} \right )^2 \,\,\,,
\eeq
where $b_0 = 0.788 [ 1 + 0.102 (B/B_0)^2 ]$ GeV/Gyr \cite{cpu1} with $B_0$ = 1 $\mu$G and $E_0 = 1$ GeV.  With our assumptions, Eq. \ref{diffusion} has a closed form solution \cite{cpu1,cpu2}.

\begin{figure*}
\scalebox{0.27}{\includegraphics{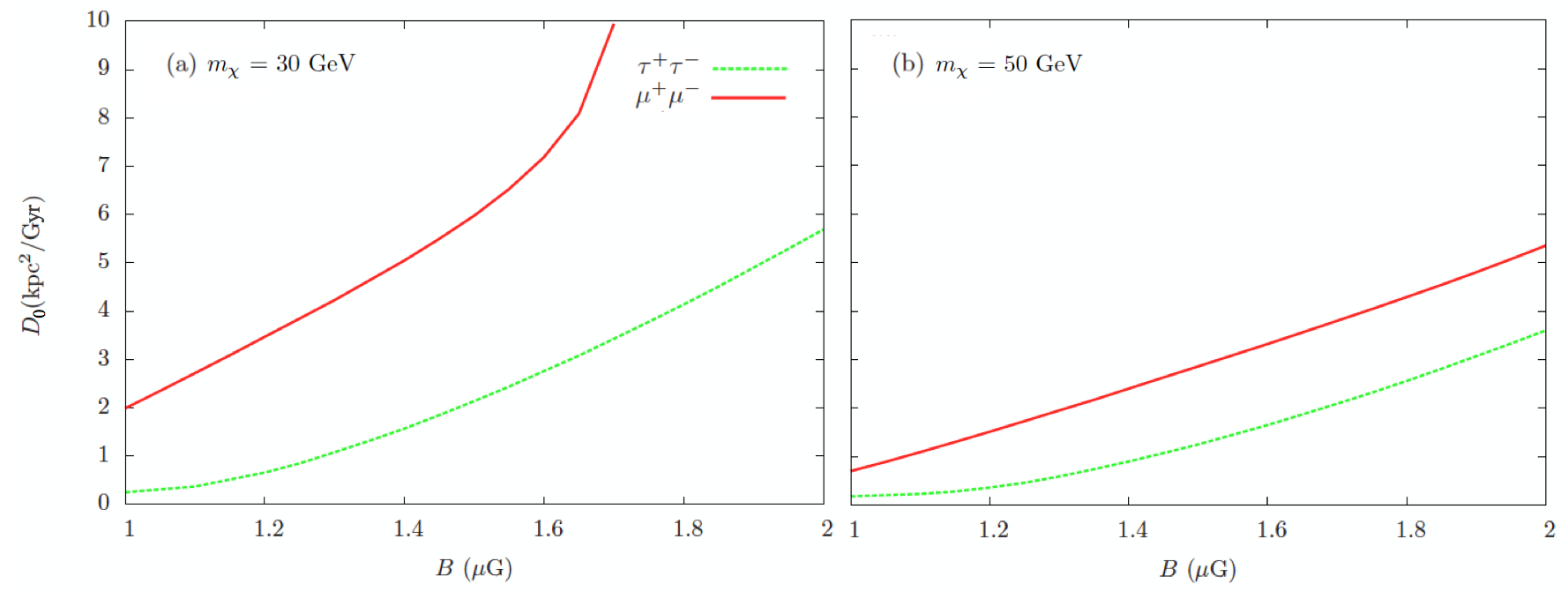}}
\caption{Exclusion curves as a function of diffusion coefficient $D_0$ and magnetic field strength $B$ for dark matter annihilation to $\mu^+\mu^-$ and $\tau^+\tau^-$ and WIMP masses $m_\chi = 30\,$GeV in plot (a) and $m_\chi = 50\,$GeV in plot (b). Models with $(B,\,D_0)$ below the curves are excluded at 95\% confidence. Our models of the $\tau^+\tau^-$ channel require only an intermediate value of $D_0$ ($D_0 \gtrsim 5.5$ kpc$^2/$Gyr) for our fiducial $B \sim 2\,\mu$G and $m_\chi \gtrsim 30\,$GeV for consistency with the data. Our most compelling limits come from the $\mu^+\mu^-$ channel: for $m_\chi = 30\,$GeV, our non-detection implies that $D_0 \gtrsim 10$ kpc$^2/$Gyr -- larger than the Milky Way value -- for $B \gtrsim 1.7\,\mu$G at 95\% confidence. 
\label{fig4} }
\end{figure*}

Not much is known about the magnetic fields of dwarf spheroidal galaxies, although our recent atomic hydrogen upper limits \cite{spekkens14} suggest that $B$ therein may be negligibly small. However, the magnetic field of the Milky Way itself, in which \seg\ is bathed by virtue of it proximity to the sun, is better understood. Authors \cite{strong_etal_2000} find a magnetic field strength $B \sim 6.1 \, \mu$G at the sun's location consistent with the radio synchrotron background at $\nu=408\,$MHz. Authors \cite{carretti_etal} used Fermi telescope observations to detect two giant radio lobes near the Galactic bulge, with fields up to 15 $\mu$G, extending up to 7 kpc from the center. Such a large magnetic field  seems inconsistent with an exponential fall off with distance from the center, as commonly assumed \cite[e.g.][]{beck09}. Authors \cite{regis_etal} have suggested a linear scaling. If this is true, then $B$ in the outer halo is substantial, and can be non-negligible at the location of the dwarf galaxies in the neighborhood of the Milky Way. Based on this linear scaling, \seg\ at $L = 23\,$kpc from the sun is bathed in a magnetic field of strength $B \sim$ 2.2 $\mu$G. We therefore consider $B = 2$ $\mu$G as a fiducial value for \seg.

The diffusion coefficient $D_0$ is more difficult to estimate for dwarf spheroidal galaxies. Authors \cite{rebusco1, rebusco2} adopted $D_0 \propto v \times l$, where $v$ and $l$ are the characteristic velocity and length scale of the gas of stochastic gas motions. From the analysis of peaked iron abundance profiles in several galaxy clusters, they found that $D_0 \sim 10^{29}$ cm$^2/$s $\sim 0.3$ kpc$^2/$Myr therein \cite{rebusco2}. This value exceeds the one for the Milky Way obtained by analyzing the ratio of Boron and Carbon isotopes ($D_{0, {\rm MW}} = 0.01$ kpc$^2/$Myr, \cite{donato_etal_2004}) by over an order of magnitude. Authors \cite{jeltema_profumo_2008} have suggested that if this scaling extends to the dwarf galaxy regime, then a comparison of the virial velocity dispersions of the Milky Way and typical ultra-faint dwarfs implies that $D_0$ in the latter systems is an order of magnitude smaller than $D_{0, {\rm MW}}$. We therefore choose our fiducial values of $D_0$ to be in the range 1 - 10 kpc$^2/$Gyr.


Once Eq. \ref{diffusion} has been solved for $dn/dE$, we may compute the local emissivity:
\beq
j_{\rm synch} (\nu,r) = \int_{m_{\rm e}}^{m_\chi} dE \,  \frac{dn}{dE} \,  P_{\rm synch}  \,\,\,.
\eeq
The synchrotron kernel $P_{\rm synch}(\nu, E)$ is given by \cite{cpu1, longair}:
\bea
P_{\rm synch} &=& \frac{\sqrt{3}}{2} r_0 e B \int_0^\pi d\theta \sin^2\theta \, F \left(\frac{x}{\sin\theta}\right ) \n
&\approx& 2.3 \; \frac{ {\rm GeV} }{ {\rm GHz} \, {\rm Gyr} } \left( \frac{B}{\mu {\rm G}} \right ) \int_0^\pi d\theta \sin^2\theta \, F \left(\frac{x}{\sin\theta}\right ), \;\;\;\; \;\;\;\;
\eea
where $r_0 = e^2/m_{\rm e}c^2$ is the classical electron radius, $x$ is given by
\beq
x \approx 0.87 \, \left( \frac{ \nu }{1.4 \, {\rm GHz}} \right ) \; \left( \frac{ \mu{\rm G} }{B} \right ) \; \left( \frac{10 \, {\rm GeV}}{E} \right )^2,
\eeq
and $F(t)$ is computed using \cite{cpu1}:
\beq
F(t) = t \int_t^\infty dz \, K_{5/3}(z) \approx 1.25 t^{1/3} e^{-t} \left[ 648 + t^2 \right ]^{1/12} \,\,\,.
\eeq
The specific intensity $\mathcal{I}$ (energy per unit time per unit area per unit frequency per unit solid angle) may then be calculated by integrating $j_{\rm synch}$ along the line of sight to the dwarf galaxy:
 \beq
\mathcal{I} (\nu) = \frac{1}{4\pi} \; \int_{\rm l.o.s.} ds \; j_{\rm synch}(\nu,s) \,\,\,.
\eeq

Fig.~\ref{fig2} shows the form of the synchrotron kernel $P_{\rm synch}$ as a function of particle energy for different observation frequencies (plot (a)), and for different values of the magnetic field (plot (b)). $P_{\rm synch}$ has a sharp cutoff at low energies which implies a corresponding cutoff at low particle masses $m_\chi = E_{\rm cutoff}$. Plot (c) shows the energy spectrum at the point of annihilation, for different channels. Direct annihilation to $e^+e^-$ results in a spectrum that is close to a delta function, with most particle at energies $\approx m_\chi$. The hadronic channel $b \bar b$ results in a much more gradual variation of the spectrum with particle energy. Plot (d) shows $dn/dE$ after transport, for $m_\chi$ = 50 GeV, $B$ = 2 $\mu$G, $D_0 = 1\,$kpc$^2/$Gyr, at position $r$ = 0.1 kpc, for the different annihilation channels.

\section{Results}
\label{results}

The mean-subtracted residual radio continuum map presented in \S\ref{data} shows no excess radio emission at the location of \seg. We now quantify the implications of this non-detection for particle dark matter properties using the models in \S\ref{models}. To maximize our ability to distinguish between the dark matter synchrotron emission and small-scale noise or artifacts, we divide the map into four $\sim1\,$degree-sized bins centered on \seg\ that contain an equal number of pixels: Bin~\#1 ($0^\circ < \theta < 1.12^\circ$),  Bin~\#2 ($1.12^\circ < \theta < 1.86^\circ$), Bin~\#3 ($1.86^\circ < \theta < 2.55^\circ$), and Bin~\#4 ($2.55^\circ < \theta < 4.0 ^\circ$), where $\theta = 0^\circ$ corresponds to the location of \seg. In order to compute the root mean square (RMS) fluctuation in intensity per bin, we use the LensTools package \cite{petri14} to simulate 1000 maps with the same power spectrum and variance as our observed map, and compute the mean intensity and RMS in each bin. 

Fig.~\ref{fig3}  shows the radiation intensity in Bin~\#1 (closest to the \seg\ stellar centroid) as a function of the WIMP mass with $B$ = 2 $\mu$G, for $D_0 = 1\,$kpc$^2/$Gyr in plot (a) and $D_0 = 10\,$kpc$^2/$Gyr in plot (b). Also shown with dashed lines is the RMS fluctuation in the data Bin~\#1 that we obtained from the simulations. We note that the $b \bar b$ channel is undetectable, since it is well below the noise level even for the optimistic choice $D_0 = 1\,$kpc$^2/$Gyr. The leptonic channels are generally easier to constrain, particularly for small $D_0$ and low $m_\chi$. The $e^+e^-$ channel predicts the largest Bin~\#1 intensity for small WIMP masses, but falls off very quickly with rising mass. This is because the energy spectrum of $e^+e^-$ due to WIMP annihilation is close to a delta function at the WIMP mass (see Fig.~\ref{fig2}c). Fig.~\ref{fig3} shows that we rule out light WIMPs that annihilate to $e^+e^-$ for $D_0 \lesssim 10\,$kpc$^2/$Gyr, a model that is also strongly disfavored by the polarized cosmic microwave background \cite{cmb_arvi}. We therefore focus on the constraints obtained for the $\tau^+\tau^-$ and $\mu^+\mu^-$ channels below.

The radiation intensity values in the four bins due to dark matter annihilation are computed for the $\tau^+\tau^-$ and $\mu^+\mu^-$ annihilation channels, and for different choices of $B$, $D_0$, and $m_\chi$. As shown by Fig.~\ref{fig3}, the synchrotron emission from dark matter annihilation in Bin~\#1 will be large and positive. Bin~\#4  is expected to be negative since the residual map in Fig.~\ref{fig1}b is mean-subtracted. Bins \#2 and \#3 may be either positive or negative depending on the annihilation channel, values of $D_0$, etc. On the other hand, in the absence of a detection we do not expect the data to show any definite trend in intensity values as we move from Bin~\#1 to Bin~\#4.

Table~1 gives the observed mean-subtracted residual map intensity values and error bars in Bins~\#1, \#2, \#3 and \#4, along with theoretical predictions for WIMP mass $m_\chi = 10$ and $30\,$GeV and for $D_0$ = 1 and 10 kpc$^2/$Gyr for the $\tau^+\tau^-$ and  $\mu^+\mu^-$ channels. The mean intensity of the data in all four bins is consistent with zero within uncertainty. As expected from Fig.~\ref{fig3}, Bin~\#1 values for the models are positive and larger for smaller WIMP masses and smaller values of $D_0$. Model Bin~\#4 values are all negative. By and large, the model values in Table~1 exceed those for the data in (absolute) amplitude in all bins, particularly for $m_\chi = 10\,$GeV; we can therefore place competitive bounds on the $\tau^+\tau^-$ and  $\mu^+\mu^-$ annihilation channel WIMP properties for $D_0 \lesssim 10 $ kpc$^2/$Gyr and $B \lesssim 2\,\mu$G.

We now determine the 95\% exclusion curves in the $D_0 - B$ plane for WIMPs of different masses annihilating into $\tau^+\tau^-$ and  $\mu^+\mu^-$ by computing the $\chi^2$ quantity for each model, summed over the four bins:
\beq
\chi^2 = \sum_{i=1}^4 \left( \frac{\mathcal{I}_{data,i} - \mathcal{I}_{model,i}}{\sigma_i}  \right )^2 \,\,\,,
\eeq
where $\sigma_i$ is the RMS fluctuation in each bin obtained from the data simulations. The likelihood function is then given by:
\beq
\mathcal{L} = \exp \left\{ -\frac{1}{2} \left( \chi^2 - \chi^2_{\rm min} \right ) \right\} \,\,\,.
\eeq
Each exclusion curve is obtained by computing the central 95\% area under the likelihood curve, assuming uniform priors. 


Fig. \ref{fig4} shows 95\% exclusion curves for WIMP annihilations into $\tau^+\tau^-$ and $\mu^+\mu^-$ for two representative WIMP masses, $m_\chi = 30\,$GeV in plot (a) and $m_\chi = 50\,$GeV in plot (b). The region below each curve is excluded by the data.  Recall that the fiducial values for the diffusion coefficient and the magnetic field strength are $D_0 = 1 - 10$ kpc$^2/$Gyr (the lower bound from scaling arguments, and the upper bound equal to the Milky Way value) and $B = 2\,\mu$G (given \seg's proximity to the Milky Way), respectively (\S\ref{models}). The exclusion limits are therefore quite strong for light WIMPs annihilating to $\mu^+\mu^-$: for $m_\chi = 30\,$GeV, our non-detection implies that $D_0 \gtrsim 10$ kpc$^2/$Gyr -- larger than the Milky Way value -- for $B \gtrsim 1.7\,\mu$G, while we exclude $m_\chi = 30\,\mathrm{GeV} \rightarrow \mu^+\mu^-$ for $B \gtrsim 1$ $\mu$G and $D_0 \lesssim 2$ kpc$^2/$Gyr.  In combination with the values of Table~1, Fig.~\ref{fig4} illustrates that WIMPs with $m_\chi \lesssim 30\,$GeV annihilating to $\mu^+\mu^-$ -- such as those favored for that channel in the dark matter interpretation of the Reticulum~II gamma-ray excess \cite{ret1} -- are ruled out by our data unless $B$ is significantly smaller (or $D_0$ is significantly larger) than argued here. 

Fig.~\ref{fig4} shows that our limits on the $\tau^+\tau^-$ channel are weaker, requiring only that $D_0 \gtrsim 5.5$ kpc$^2/$Gyr for $B \sim 2\,\mu$G and $m_\chi \gtrsim 30\,$GeV for consistency with the data. It is therefore unlikely that we would detect the $m_\chi = 30\,\mathrm{GeV} \rightarrow \tau^+\tau^-$ signature if $D_0 \sim 10$ kpc$^2/$Gyr, as might be expected if the synchrotron radiation in \seg\ is generated by the magnetic field of the Milky Way in which it is bathed.

\section{Conclusions}

In this article, we placed competitive bounds on dark matter properties using radio observations of the ultra-faint dwarf spheroidal galaxy \seg. This target is particularly compelling for radio synchrotron dark matter searches, because its close proximity ($L \sim 23\,$kpc) implies that it could be bathed in a non-negligible halo magnetic field of the Milky Way. \S\ref{data} presented a $\nu = 1.4\,$GHz map of a $4^\circ \times 4^\circ$ region around \seg\ obtained with the GBT. Point sources were subtracted from the map using the NVSS catalog, and the diffuse sky was subtracted using a Planck map. Our non-detection near the stellar centroid of \seg\ in the resulting mean-subtracted residual map was used to place bounds on the WIMP mass for different values of magnetic field $B$ and the diffusion coefficient $D_0$.

\S\ref{models} presented our models for the radio synchrotron emission expected from \seg\ for different dark matter annihilation channels, adopting an Einasto density profile for the dwarf and matching its Fermi emission measure. We studied the diffusion of charged particles in a magnetic field, computing the annihilation rate after transport determined by $D_0$. The synchrotron radiation at a given observing frequency is sensitive to $B$, $m_\chi$ and the annihilation channel. 

\S\ref{results} used these models to place bounds on particle dark matter properties from our observations. Since the dark matter synchrotron intensity has very little power on small scales, we binned our data into four $\sim1\,$degree-sized bins in order to maximize the dark matter signal while minimizing contributions from thermal noise and small-scale artifacts. We computed the likelihood function for different WIMP masses, and placed bounds on dark matter properties. Our data strongly rule out light WIMPs annihilating to $e^+e^-$, but are not sensitive to the $b \bar{b}$ channel. Our models of the $\tau^+\tau^-$ channel require only an intermediate value of $D_0$ ($D_0 \gtrsim 5.5$ kpc$^2/$Gyr) for our fiducial $B \sim 2\,\mu$G and $m_\chi \gtrsim 30\,$GeV for consistency with the data. Our most compelling limits come from the $\mu^+\mu^-$ channel: for $m_\chi = 30\,$GeV, our non-detection implies that $D_0 \gtrsim 10$ kpc$^2/$Gyr -- larger than the Milky Way value -- for $B \gtrsim 1.7\,\mu$G at 95\% confidence. This limit is inconsistent with the $m_\chi \lesssim 10\,$GeV$\,\rightarrow \mu^+\mu^-$ interpretation of the Reticulum~II gamma-ray excess unless $B$ is significantly smaller than the fiducial value adopted here. 

\seg\ is a unique target for radio synchrotron dark matter searches because of its close proximity to the Milky Way, but a joint radio analysis of several promising dwarf spheroidal galaxies could also place powerful constraints on particle dark matter. We have obtained $\sim$300 hours of wide-field GBT mapping observations in the vicinity of 6 such systems in order to do just that. A detailed analysis of this data is left to future work.

\acknowledgments{We thank Beth Willman, Judith Irwin, and Lawrence Widrow for helpful suggestions, and acknowledge assistance in the data reduction from Bryan Fichera and Melissa Diamond. AN is grateful for financial support from Queen's University, and acknowledges hospitality of the University of Pennsylvania. KS acknowledges support from the Natural Sciences and Engineering Research Council of Canada.  The National Radio Astronomy Observatory is a facility of the National Science Foundation operated under cooperative agreement by Associated Universities, Inc. }

\bibliography{references}

\end{document}